# DOMAINS OF STATES OF CHEMICAL SYSTEMS:
# LE CHATELIER RESPONSE, STRUCTURE OF THE DOMAINS AND EVOLUTION

B. Zilbergleyt

System Dynamics Research Foundation, Chicago, USA
livent@ameritech.net.

Previous publication [1] has briefly discussed the idea of Le Chatelier Response (LCR) as a measure of what equilibrium chemical system undertakes to relieve stress induced by an external impact. The objectives of this work are to investigate further on the influence of the LCR on the system behavior under stress, the shape of its domains of states in terms of traditional and dynamic bifurcation diagrams, and the system proneness to evolution. The usage of maps in thermodynamics of the chemical systems is discussed. Thermodynamics of a chemical trigger, designed in similarity with laser, is described. Such a development is very important in a context of new model of chemical equilibrium; on the other hand, it may motivate and facilitate a progress in other applications of the LCR and the Method of Bound Affinity [2][i] to the chemical and alike systems.

## INTRODUCTION
Though the term evolution is being vaguely applied to a variety of processes, with regard to the chemical systems we will mean by it only development based on choice by chance, potentially leading to a new quality. Well argued similar point of view is discussed in [3]. We will be using throughout this paper the pitchfork bifurcation as a major example. Because equilibrium state is the only state within the capacity of thermodynamics, both terms - state and equilibrium state - are actually equivalent. The first thing to keep in mind is that chemical equilibrium coincides with *true thermodynamic equilibrium* (TDE) only in isolated chemical system. In compliance with common sense and tradition to talk of the system states as "at" or "far-from" equilibrium [4], we evaluate general state of the system by deviation of its chemical equilibrium from its *true thermodynamic equilibrium*. In our theory, reaction *shift from equilibrium*, a distance from the system state to TDE, in terms of reaction extent is $\delta\xi=1-\Delta\xi$, or just $\delta=1-\Delta$ (clearly $\Delta=1-\delta$), where $\Delta\xi$ is the reaction extent; it takes on unity at TDE [5]. To qualify the system interaction with its environment numerically and to account for the system internal intricacy, we have introduced the Le Chatelier response (LCR) as an explicit function of the system shift [1]. In this work we will use a symbolic reaction A+B=cC (reaction R1) or PCl3+Cl2=PCl5 (reaction PCl5); if the stoichiometry is (-1,-1,1), R1 matches the $PCl_5$ reaction. All the numerical solutions were obtained for isobaric-isothermic conditions and Gibbs' free energy as characteristic function.

The most habitual value to depict the reaction strength is $\Delta G^0$; we employ more informative *thermodynamic equivalent of chemical transformation* $\eta$ − mole number of the reaction participant, transformed in the chemical reaction from its initial state to TDE, per its stoichiometric unit (explained in [5]). Being a function of $\Delta G^0$ and initial reactants mole ratio, the thermodynamic equivalent of transformation reflects the chemical system ability to withstand perturbations of equilibrium, and may be obtained by calculation of equilibrium composition in isolated mono-reaction system. It depends unambiguously on initial reactants mole ratio and temperature, that is on $\Delta G^0$, as illustrated by Fig.1.

---

[i] In some publications we tentatively called it the method of chemical dynamics for using explicit thermodynamic forces (TDF); the bound affinity is an implicit force.

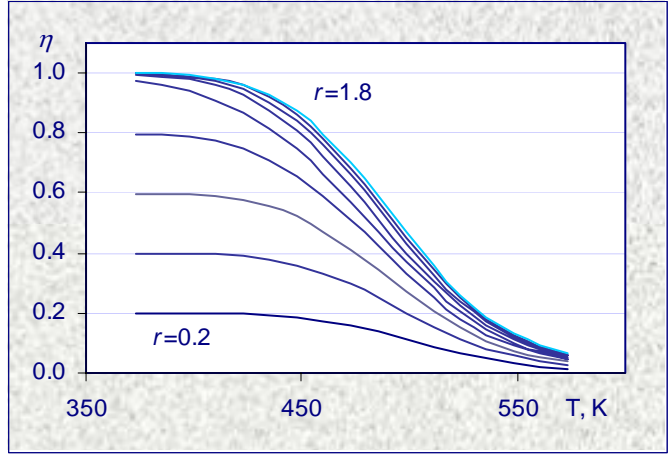

Fig.1. Dependence of the thermodynamic equivalent of transformation $\eta$ on temperature and initial reactants ratio $r$, $PCl_5$. The curves correspond to the initial amount of $PCl_3$ of 1 mole and varying $Cl_2 : PCl_3$ mole ratio from 0.2 to 1.8 with the step of 0.2.

The same dependence is typical for many reactions. In a particular case in Fig.1, $\Delta G^0$ of reaction increases with temperature; the curves converge to low values, and equilibrium point is achieved at less consumption of the reactants.

THE LCR AND DERIVATION OF THE BASIC LOGISTIC MAP

We will derive the chemical system basic equation of state, taking a complicated form of the LCR [1] instead of accepted earlier on equating it merely to reaction shift [5]. By definition, any open system is at the same time a subsystem, and at chemical equilibrium the speed of a chemical reaction in the subsystem, expressed as a function of the internal $A_{ji}$ and external $A_{je}$ thermodynamic forces, is

(1) $\qquad\qquad\qquad\qquad\qquad\qquad\qquad A_{ji} + \Sigma o_j A_{je} = 0,$

$o_j = a_{je}/a_{ji}$ is reduced (generalized) Onsager coefficient. The internal thermodynamic forces are De Donder's thermodynamic affinities (or *eugenaffinities*) [6]

(2) $\qquad\qquad\qquad\qquad\qquad\qquad\qquad A_{ji} = - (\delta\Phi_j/\delta\xi_j)_{x,y},$

$\Phi_j$ stands for major characteristic functions of the system or its enthalpy at corresponding thermodynamic parameters x,y. External generalized thermodynamic force $A_{je}$ represents external impact on the open chemical system, and at chemical equilibrium is mirrored by the bound affinity; that's why we call the entire method and related theoretical model the *method of bound affinity* [2]. Using the LCR of the j-system as

(3) $\qquad\qquad\qquad\qquad\qquad\qquad\qquad \rho_j = \Sigma\nu_p\delta^p,$

the weights $\nu_p$ are unknown and therefore are to be equalized to unities. On the other hand, we can express the system response via external thermodynamic force

(4) $\qquad\qquad\qquad\qquad\qquad\qquad\qquad \rho_j = - (1/\alpha_j)\Sigma o_j A_{je};$

$\alpha_j$ is just a coefficient, the minus sign means that the system is changing its state to decrease impact of the external TDF. The LCR as a function of the reaction shift is dimensionless, dimension of $A_{je}$ is energy[i]; $O_j$ as a relative value has no dimension, therefore the $\alpha_j$ dimension should be energy. Combining expressions (3) and (4) to express the thermodynamic forces, substituting expression (2) and these new expressions to equation (1) and multiplying its both sides by $\Delta_j$, we obtain a condition of chemical equilibrium in open chemical system as

$$-\Delta\Phi_j(\eta_j,\delta_j)_{x,y} - \alpha_j\,\rho_j\Delta_j = 0. \quad (5)$$

Let the characteristic function be Gibbs' free energy; after reducing both terms by RT and introducing *reduced chaotic temperature* $\tau_j = \alpha_j/RT$, we obtain intermediate expression

$$\ln[\Pi_j(\eta_j,0)/\Pi_j(\eta_j,\delta_j)] - \tau_j\varphi(\delta_j,p) = 0. \quad (6)$$

The value of $\Pi_j$ is traditional mole fractions product: the numerator under the logarithm corresponds to TDE, or $\delta_j = 0$ and therefore equals to constant of equilibrium $K_j$, the denominator is related to a state, deviated from TDE by $\delta_j \neq 0$. Factor p is the largest power of the reaction shift in the LCR. Expression (6) is a generalized logistic map

$$f(\delta_j) = \tau_j\varphi(\delta_j,p). \quad (7)$$

We would like to restate that map (6), by analogy with the Verhulst's model of the population growth [7], includes shift $\delta_j$ as a parameter of state, $\tau_j$ as "growth" parameter, and p as the LCR parameter; $\Pi_j(\eta_{kj},0)/\Pi_j(\eta_{kj},\delta_j)$ is a reverse value of relative "*chemical population*" factor. The numerator carries information of TDE, or the maximally achievable population size at given initial conditions and thermodynamic variables, representing carrying capacity of the system. Parameter $\tau_j$ defines "growth" of the deviation from TDE; like in the Verhulst's model, it is a fraction where the numerator is an "energy" equivalence of the external impact on the system (the "demand for prey" in the "prey-predator" model in bio-populations [8]), while the denominator (RT, the system "thermal energy") is a measure of the system capability to resist the changes of its state.

The value of $\delta_j$ is supposed to fall into interval [-1,+1]. Solutions to the map (6) are essentially different depending on whether the start value p-set is unity or zero (we will mark them U-set and Z-set correspondingly). The start factor splits the solutions by 2 groups that seem to be relevant to quite different types of chemical systems. The first case looks more common, and we will begin with it. Applying the method of mathematical induction to U-set, one can easily get

$$\varphi_u(\delta_j,p) = \delta_j(1-\delta_j^p), \quad (8)$$

subscript at $\varphi$ shows the choice of the p-set. Substituting (8) to (6), we obtain

$$\ln[\Pi_j(\eta_j,0)/\Pi_j(\eta_j,\delta_j)] - \tau_j\delta_j(1-\delta_j^p) = 0. \quad (9)$$

At this point we reached to the point to distinguish between the chemical reaction values and the chemical system values. The first term of map (9) is the change of Gibbs' free energy in j-reaction; in case of isolated j-system this value is relevant to the reaction. The second term reflects interactions between j-subsystem and its environment (or its compliment to a bigger system) and definitely represents a system value. Together they comprise the left side of map (9), *that is nothing else but full change of Gibbs' free energy in open chemical system, that is definitely a system value. Being equated to zero it*

---

[i] More exactly, it is an energy equivalence for the derivative of the external impact with respect to dimensionless reaction extent.

*defines a state of open chemical system. At $\delta_j=0$ it turns into classical change of Gibbs' free energy for isolated chemical system where the ideas of the reaction $\Delta G$ and the system $\Delta G$ become undistinguishable.*

Simplified version of map (9) at p=1, or conventional logistic map, was obtained in [5]. Map (9) describes development of chemical populations at certain thermodynamic conditions: it maps population of j-system at TDE onto population of the same system, whose state was forcefully shifted by external impact to another state of chemical equilibrium, where the external force is balanced by the bound affinity. Its graphical solutions, known as bifurcation diagrams, constitute domain of states of the chemical system. It is remarkable that we obtained map (9) exclusively from known or introduced (like the LCR) thermodynamic ideas, not using any criteria of populations and their growth at all. The results show that there is a certain similarity between biological and chemical populations.

Traditional bifurcation diagrams – solutions to simple classical logistic maps like

(10) $$x_{n+1}= \lambda x_n(1-x_n)$$

with relative population size $x_n$ (or $x_n \sim \Delta_j$ in our terms) [9]. All key relations between the population and its environment in a competition for resources as well as predatory factors are immanent for the growth factor $\lambda$. Ironically, only the area of bifurcation diagram beyond the split point ($\lambda \gtrsim 3.1$), where bifurcations occur at the first time and then lead to proliferation of branches is a usual subject to detailed discussion. In a traditional approach the part of the curve at $\lambda \leq 1$ resting on abscissa is thought to be the area of the population extinction. The ascending part between the extinction area and the first bifurcation point seems to have no meaning at all, being merely a continuation of thermodynamic branch.

From the very beginning of this project (back in 1990s) the author has used another type of diagrams, with the system shift from equilibrium instead of reaction extent versus reduced chaotic temperature; one can call them *inverse diagrams*. Besides that, as we will show further, traditional bifurcation diagram is not enough informative and, besides that, doesn't provide for a good visual image of the chemical system domain of states. To amend it, later in this paper we will introduce a variation of the diagram in terms of shift vs. thermodynamic force that we call a *dynamic inverse bifurcation diagram*. The old fashioned diagram will be distinguished as *static inverse bifurcation diagram*. Each diagram reveals different features of the chemical systems, and both are useful as complementary. If this work we use only the inverse diagrams and domains of states.

STATIC DOMAIN OF STATES OF THE CHEMICAL SYSTEM

Though being obtained purely from the thermodynamics' laws and logic, map (9) leads to solutions that are virtually identical to traditional bifurcation diagrams, featuring the same areas and break points. Bifurcation diagrams constitute the chemical system domain of states as shown in Fig.2. The more is the reaction robustness, or the larger is the $\eta$-value, the larger is its TdE area limit. Varying p splits $\eta$-stems by various branches of the same general shape but with different branch stemming up parameters, $\delta$ and $\tau$, as shown in Fig3. In the static inverse bifurcation diagram, the area with $\delta_j=0$ belongs to TDE by definition: any perturbation of equilibrium within this area will extinct; the state is defined by the classical part of map (9) because its second term turns to zero. The system is protected against external impact and behaves itself as isolated. It is obvious from Fig.3 that the values of p don't affect the TdE area limit. We call next part of the diagram the open equilibrium area, or OpE. Within this area equilibrium in the open system is defined by full map (9); it can essentially deviate from TDE up to the δ value at the bifurcation point. This area ends up with the first bifurcation point. Then follows bifurcation area, its meaning is quite the same as in traditional case with one interesting difference. Because domain of states in Fig.2 is precisely a domain of chemical equilibrium states, the pitchfork branches are inevitably the attractors and simultaneously the limits for the amplitudes of potential chemical oscillations in the system, confined within bifurcation area by the external force. Such an understanding of the areas of bifurcation diagrams is very important for simulation of complex chemical equilibria. Though the simulation program that we used to obtain the above results was adjusted to stop by the end of period-2 bifurcations, one should expect similarities also beyond that area.

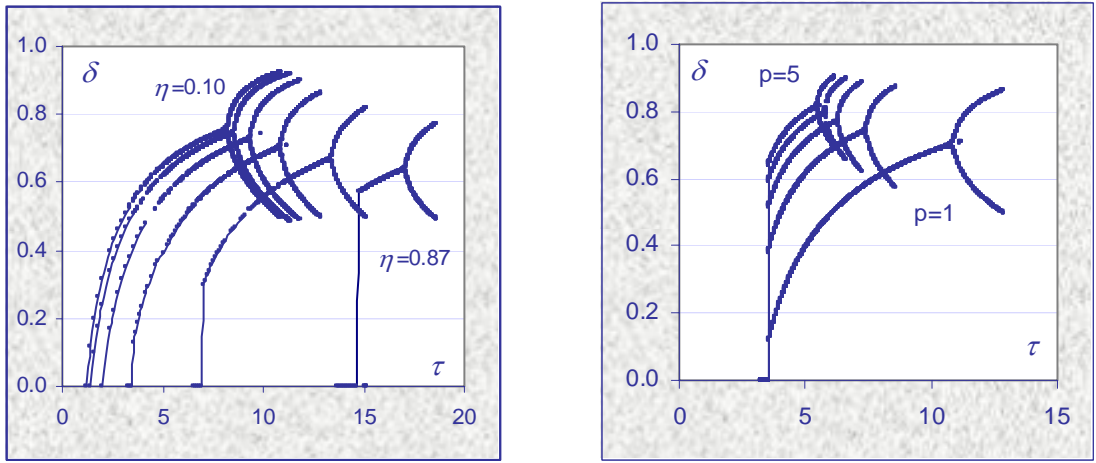

Fig.2, left: Static domain of states of the chemical system, R1, p=1, varying η.
Fig.3, right: A "singular" static bifurcation diagram for the same reaction, fixed η, varying p. The curves stem from the same basic branch at η=0.57, p-step=1.

Besides the iteration method, there is also an analytical opportunity to find the TdE area limit [5]. Analytically found data showed an excellent match with the simulation results at the most part of the η values from its [0,1] interval.

Along δ−axis the static domain of states is restricted by unity – no one chemical reaction cannot be pushed back below its initial state, so $δ_j ≤ 1$. As concerns to the τ-axis, we do not know whether this limit exists at all and how important its existence and value might be. So far the model we are developing didn't need such an idea as the τ-limit.

Interestingly enough that traditional logistic map is easily convertible to deviations from equilibrium, or stationary regular state in case of bio-populations. Indeed, the factor $(1−x_n)$ of the map (10) means deviation of the livestock amount from stationary regular value, similar to the chemical system shift from equilibrium. It allows us to convert map (10) from relative population to population shift, maintaining exactly the same shape of the map but in terms of $δ_n=(1−x_n)$ instead of $x_n$. Such an inversion may put new interesting contents into the static diagram areas in application to bio-societies.

DYNAMIC BIFURCATION DIAGRAMS AND DYNAMIC DOMAIN OF STATES

Traditional bifurcation diagrams give a good idea of what happens to the system with increase of the control parameter, $\lambda$, $\tau_j$ or alike, covering all possible scenarios. In our opinion, such a description is too abstract in case of chemical systems, and contains more mathematical than physico-chemical information which is not enough comprehensive. Indeed, any response should be tied with a reason: in case of chemical system the reason is thermodynamic force while corresponding shift from TDE (or a function of it) is the response. This way we get to the idea of bifurcation diagram in the force-shift terms, or *dynamic bifurcation diagram,* pictured in following 3 figures. The obvious expression for TDF via change of Gibbs' free energy in finite differences is $f_j = -\Delta G_j / \Delta_j$. Dividing (5) by

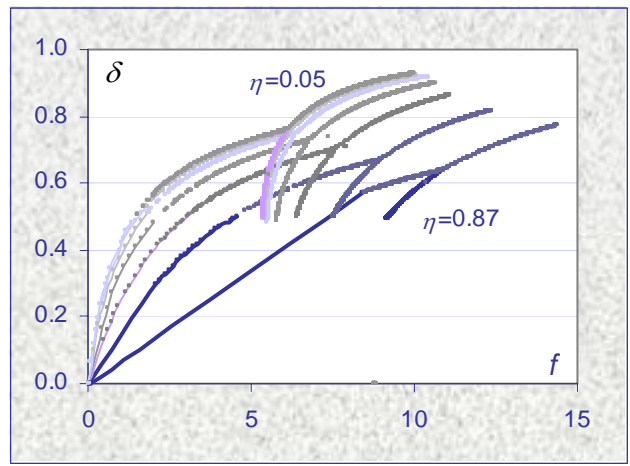

Fig.4. Solutions to the map (12), dynamic bifurcation diagrams, R1, p=2, varying $\eta$.

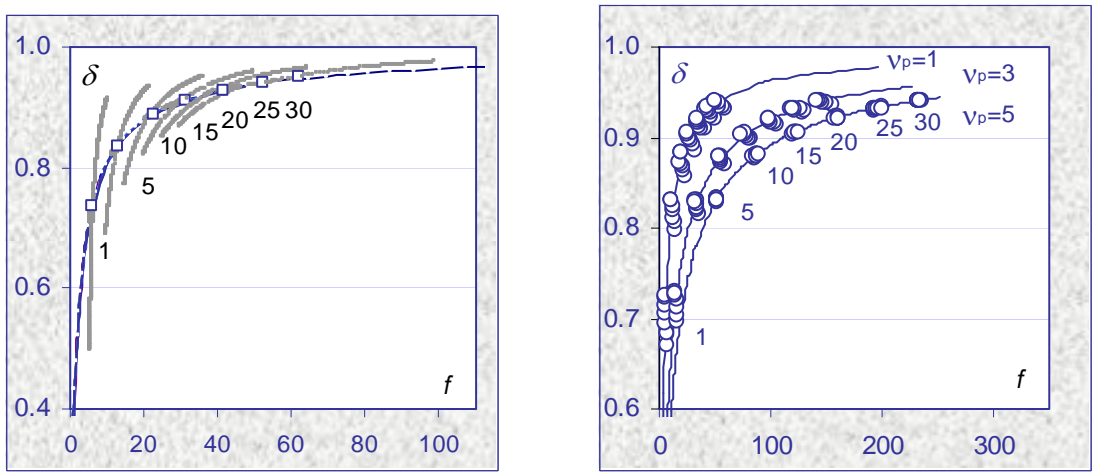

Solutions to the map (12), dynamic bifurcation diagrams.

Figs.5, left: The trunk line with growing out bifurcation wings attached approximately at the square white points, R1, $\eta$=0.20, numbers at the branches are p-values.

Fig.6, right: The "evolution trunks" with the clusters of bifurcation points sitting on them.

$\Delta_j$ one can get alternative expression of the external TDF, reduced by RT

(11) $\qquad\qquad\qquad\qquad\qquad\qquad f_{uj}=\tau_j\delta_j(1-\delta_j^{\,p})/(1-\delta_j),$

and map (9) now may be re-written as a balance of thermodynamic forces

(12) $\qquad\qquad\qquad\qquad [1/(1-\delta_j)]\ln[\Pi_j(\eta_j,0)/\Pi_j(\eta_j,\delta_j)] - f_j = 0.$

The locuci of the bifurcation points, the *trunks*, correspond to various sums of stoichiometric coefficients $\Sigma\nu_p$ of the reaction products. All the trunks drop sharply to zero point of coordinates. In a simple case of p=1 map (12) turns into

(13) $\qquad\qquad\qquad\qquad [1/(1-\delta_j)]\ln[\Pi_j(\eta_j,0)/\Pi_j(\eta_j,\delta_j)] - \tau_j\delta_j = 0.$

The first term in this expression represents the *bound affinity* [2].

As examples of the external TDF one may tell a chemical impact from a parallel chemical reaction, linked to the system by common participants; an electrical force in electrochemical system where the shift from equilibrium is supported by electrical potential (like in electrically inhibited corrosion); a force moving the system from its state in case of change in temperature of equilibrium system.

Fig.4 shows non-traditional bifurcation diagram in $\delta$–$f$ coordinates (compare to Fig.2). Dynamic diagrams skip the TDE area and the OpE area stems out immediately from zero point of the reference frame . It is clear from Fig.4 why the magnitude of $\eta$ shows the robustness of chemical system – the less is $\eta$ the bigger is the system shift from TDE under the impact of the same external force.

Being re-plotted for fixed $\eta$–values and varying p, dynamic bifurcation branches undergo interesting transformations (Fig.5). First, all the period-2 bifurcation points, corresponding to the same $\eta$, fall onto unique common curve (a trunk), along which the reaction shift value asymptotically tends to unity as the p and f values increase. All bifurcation points are residing on the *evolution trunks* individually, in Fig.6 in clusters. The branches between their period-2 bifurcation points and period-4 points in Fig.5 look like wings, increasing f-value prompts the wings to rotate gradually, asymptotically folding towards the evolution trunk and eventually merging it. The clusters in Fig.6 are comprised from closely sitting bifurcation points for various $\eta$ (0.05,…,0.97). Lines, connecting different points within the clusters are rotating similarly to bifurcation branches. Obviously, the higher is p the harder is to achieve corresponding bifurcation point, and at large p, where the wings fold down to the trunk, most probably no visible evolution will occur. Like the static bifurcation diagram and on the same reason, the dynamic diagram is restricted along the $\delta$–axis by unity.

## WHY THE MAP?

Several steps allowed us to develop a unified model of chemical equilibrium that covers the entire spectrum of thermodynamic states, from true equilibrium to true chaos. Those steps were: the strong attachment of chemical reaction to a certain chemical system; re-writing the thermodynamic functions in terms of the chemical system shift $\delta_j$ from TDE; equation for the Le Chatelier response; and discrete technology. It is hard to argue with the first step; most of the manuals in chemical thermodynamics persistently prove very poor system thinking, failing with the basic definition of thermodynamic system as the major object of that science. The second step leads to exact quantitative definition of the system operation areas instead of their description in linguistic variables like "close" or "far" from TDE [9]. One may accept the LCR in principle, or reject the way we introduced it; perhaps less arguable is simplified map at p=1. As concerns to the last of the above listed points, one should recall that all natural objects as well as the models, that science is using to describe them, are continuous and discrete as well, depending on size of the measuring stick we use to quantify space and time, in great similarity with famous Mandelbrot's task of the sea shore length [10]. Also, we may add that chemical thermodynamics always used - and in visible future will be using - the delta-values like $\Delta H$, $\Delta S$, etc. on the well-known reasons; among them is the fact that only differentials of thermodynamic functions are available from experiment [11]. The delta-mentality is a welcome part of this science.

In classical theory the isolated system is a reigning routine, closed system is exclusion, and the open system in most cases barely proceeds beyond a declaration of its openness. That's why classical theory uses continuous equations and operates exclusively with TDE. If any new theory is trying to unite both chemical thermodynamics in one, usage of maps looks inevitable. Our results have proven it: map (9) reserves space for open and – at the limit end - for isolated chemical systems as well. The characteristic function becomes non-differentiable and a demand for discreteness arises as soon as the second term of the maps (9) and (12) turns to a non-zero value.

Though the map like (10) transforms n-subscribed value into (n+1), both sides of map (9) are subscribed identically. The reason is simple – the whole theory is built up around chemical equilibrium, and the iteration process ends when $\delta_{n+1} - \delta_n < \varepsilon$, at the point where both values cannot be distinguished at given measure of accuracy. This is an exact mirroring of how the real system approaches the equilibrium. On the other hand, as it was already mentioned, map (9) plots population of one state onto population of the same system at another state.

EVOLUTION

The advantage of dynamic bifurcation diagrams is a great visual power to communicate evolution of the chemical system. In Fig.6 one can see how the bifurcation points are planted on the evolution trunk, taking their places depending upon the LCR power series. Accepting that evolution occurrence is strictly a function of the bifurcation point

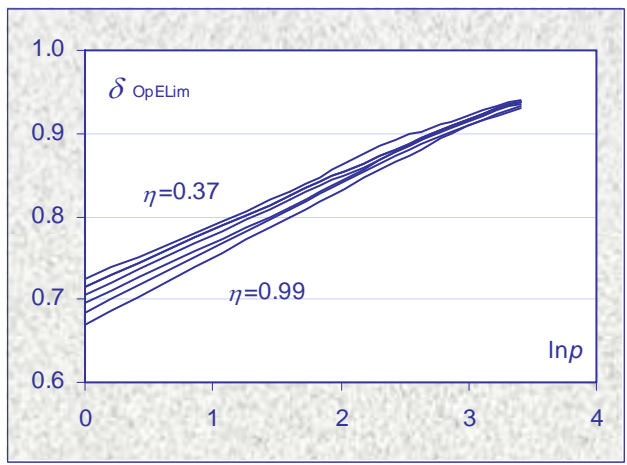

Fig.7. Shift value at the bifurcation point vs. logarithm p, R1, varying $\eta$.

attainability, two observations may be done:

1.  The more complex is the system response (or the larger is p) the larger should be the external force to bring the system to the evolution point. In other words, *the more complex is the system response the less is the system proneness to evolution*. From the energetic point of view, it is important what amount of energy has to be applied to the chemical system in order to push chemical reaction back towards the initial state.

2.  The less the system is prone to evolution the larger shift value must be achieved before bifurcation point.

In addition to Fig.6 we have found very interesting dependence of the shift value at the bifurcation point upon the p value (Fig.7). Obvious linearity of $\delta_{OpELim}$ vs. ln$p$ is amazing; at the present moment we cannot explain this effect.

The product $(\tau\delta)_{OpELim}$, being equal to thermodynamic force to be applied to the chemical system in order to achieve bifurcation point, corresponds to the dynamic, and $(\tau\delta\Delta)_{OpELim}$ − to the energetic thresholds of the thermodynamic branch stability, followed by the break of symmetry and formation of the bifurcation branches. This observation is very important for explanation of the energetic background of bifurcation phenomena.

## SPECIAL CASE: Z-SET BASED SOLUTIONS AND CHEMICAL TRIGGER

As it was mentioned above, Z-set leads to very different basic expression. It can be easily shown that in this case the $\varphi$−function is

$$(14) \qquad\qquad\qquad\qquad \varphi_{zp}(\delta_j,p)=(1-\delta_j^p),$$

and brings following expressions

$$(15) \qquad\qquad\qquad ln[\Pi_j(\eta_j,\,0)/\Pi_j(\eta_j,\,\delta_j)] - \tau_j(1-\delta_j^p)= 0,$$

for the basic map, and

$$(16) \qquad\qquad\qquad\qquad f_{zj}=\tau_j(1-\delta_j^p)/(1-\delta_j)$$

as force. The small differences between similar expressions lead to essential differences in the domains of states, shown for Z-set case in Fig.8 and Fig.9. The major feature of the static diagrams with Z-set is that the curves start immediately with the OpE area. The second common feature of the solutions is that beyond bifurcation points the separated branches literally hurry up to the limit values of zero or unity. In case of U-set upper branches are tending to unity asymptotically and at low η-values quite slowly, while positions of the lower branches cannot be foretold. Both static and dynamic bifurcation Z-diagrams are very similar. The dynamic domain of states for R1 with fixed value of η=0.37 is shown in Fig.10. One can see the evolution trunk with highlighted approximate positions of bifurcation points. The diagram with p=0 is shown in Fig.10 just to show that it doesn't experience any bifurcation at all.

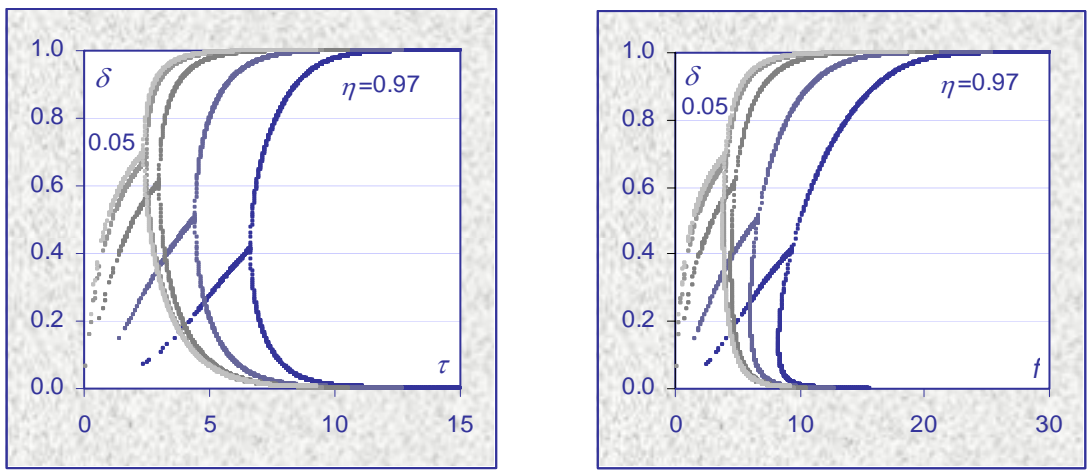

Fig.8. (left) Static and Fig.9. (right) Dynamic domains of states of the chemical trigger, Z-set, R1, p=2.

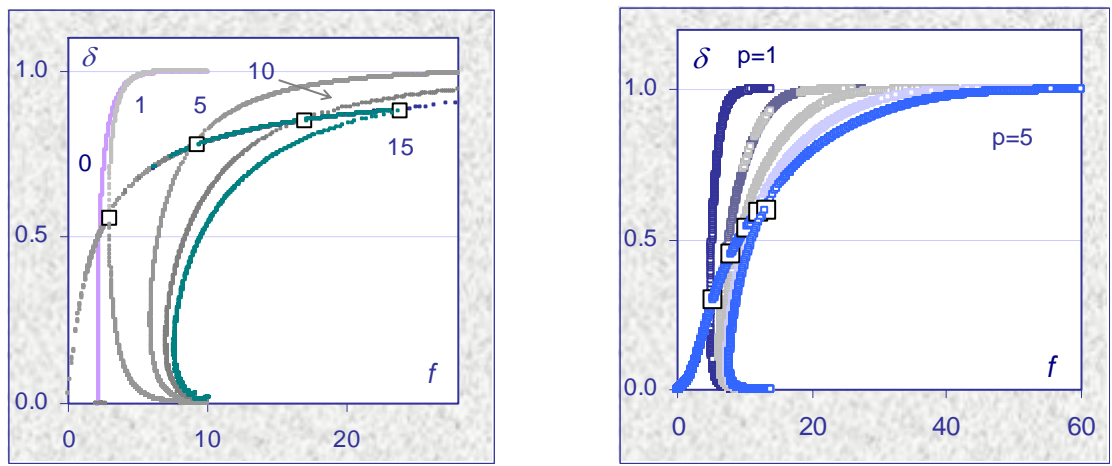

Fig.10, left.  Dynamic domain of states, R1, η=0.37 (Z-set), numbers show the p-values.
Fig.11, right. Dynamic domain of states of a chemical trigger designed as laser, η=0.99.

The shape of bifurcation Z-diagrams instigates a guess that they and transformations behind them belong to a group of triggering processes. The simplest chemical trigger may be designed like a laser with basic chemical reaction M+*f*=M*, where M is represents the particles to be activated to M* and *f* is the pumping force. Indeed, the simplest laser has 2 levels and can be illustrated by Fig.11 showing its dynamic domain of states at Z-set. The δ value equals to the ratio M*/(M+M*) – the closer it is to unity the more particles are pumped to the highest level. In thermodynamic equilibrium, in absence of the pumping force nearly all of the residents are M, sitting on the ground level, δ=0. In our terminology, for such a system η≈1 (in Fig.11 we set η up to 0.99). Driven by the pumping force, particles get excited M→M*, and jump to the pump level; M-population decreases, and eventually we may achieve the inverse population with δ≈1. Beginning at *f*=0 and up to bifurcation point, shift from TDE follows the OpE curve. All δ-*f* curves in the OpE areas for any p are resting exactly on the evolution trunk; *the evolution trunk may be compared to the line of initial pumping*. At the bifurcation point, which is akin to the laser threshold (see [12],[13]), a bistability occurs due to almost jump wise (very clear at p=1!) splitting of the branches between δ=0 and δ=1. In this area the ground and the activated levels are both stable: pumping force, i.e. the external TDF in our terms, moves the population to the pump level while spontaneous collective irradiation moves it back to ground level.. The p-set order, or the highest p-value in the p-set, may be roughly identified with the number of possible transition paths from  the pump to ground level; that's why Z-set {0} doesn't make any sense. In case of 4 energy levels, typical for contemporary laser devices, maximal p-value in the set is 5. In a particular couple of levels (i, i+1), one of them may be the pump or the ground one, we have p=1, and a wide pitchfork bifurcation follows the bifurcation point (Fig.11).

The bifurcation area period-2 is the terrain where the oscillatory processes – irradiation, auto-oscillations, etc. occur, but one cannot exclude that the opportunities may exist beyond further splits.

Application of Z-set based formulae is justified in full for the systems with discrete population levels (0,1) (imaginary speaking, "live or die" systems), or in general to the systems working in turn-on – turn-off modes. Zero power of the shift in the LCR formula with Z-set provides for a solution with guaranteed at least $\rho=1$ (that is equal to the shift at the pumping level). The above picture certainly doesn't cover all possible types of triggering. Description of the process in lasers as a big pitchfork bifurcation doesn't exclude occurrence of the same and some other types of bifurcations, found in real devices.

CONCLUSION

We would like to remind the reader that all premises of the theory, derivations and formulas in this paper are relevant to chemical equilibrium.

All of the contents of this paper can not be applied to the flying in air chemical reactions (as in the most of manuals on chemical thermodynamics); it is relevant to chemical systems, whose behavior is a function of their internal potential, their internal complexity that defines their response to an external impact, and the impact itself. All those three components are present in the basic map (9), first two explicitly. The complexity is defined not by the number of the system components, but by the amount and features of interactions between them. In other words, the system is complex only if it can be divided by subsystems that live different life patterns and interact each with other.

In our theory, the distinction between isolated, closed and open systems vanishes – in the contrary to classical theory, their subordination is principally turned upside down: the general case is the open system, everything else are particularities (if not to say details). The potential simulation program will be well trained in when to use the second term of the maps in concrete cases of chemical systems.

The form of the LCR, we introduced in [1] and have developed further on in this work in attempt to account for the system complexity, is definitely one of possible choices. To the best of the author's knowledge this idea is quite fresh and no experimental data to compare are available.

The objective of this work was to show interdependence between the three factors, named in the header – LCR, the system domain structure and its proneness to evolution via bifurcations. Actually this paper carries one very important message more. While all previous works in the field were focused on the *ad hoc* models (e.g. well-known and well-designed toys like "brusselator", "oregonator") and on well known but particular chemical reactions similar to BZ, we state that *thermodynamic openness inevitably leads to bifurcation diagrams for any chemical system as building blocks of its domain of states*. Most of zillions of studied up to now chemical systems often did not reach as far as to bifurcations, either due to very negative $\Delta G^0$ or $\eta \rightarrow 1$ values of reaction – or to large p-values. However, one may admit that relevant, let be scanty results in chemical practice might have been either rejected as mismatching the traditional perception or just overlooked. As opposite to BZ-reaction performing attenuated oscillations between two distinguishably colored states, one can imagine that the system states, corresponding to the opposite points of the pair of bifurcation branches, may differ only by the shift magnitude. Coexisting in the same system, being visually or otherwise indiscernible, they may be segregated spatially. The observer can measure only a weighted mean of $\delta$.

The features of equilibrium in complex chemical systems fall in the OpE area. This is quite visible on the dynamic bifurcation diagrams, where the system deviation from TDE starts at zero point of the reference frame. In certain and easily predictable cases, accounting for the non-classical term of map (9) causes tangible difference in calculated compositions, moving them much closer to real numbers; some results are given in [14]. Dynamic Inverse Bifurcation Diagrams raise the model to the next level of abstraction. They show that in the simplistic systems with p=0 evolution cannot occur (see Fig.10). With regard to chemical systems, this important conclusion may be re-phrased as that the more intricacy the system possesses the better is its non-evolutionary adaptation and less are the chances for evolution.

REFERENCES.